\documentclass{aa}  
\usepackage[draft]{todonotes}
\usepackage[colorlinks=true]{hyperref}
\hypersetup{linkcolor=blue, citecolor=blue}
\usepackage{graphicx}
\usepackage{txfonts}
\usepackage{bm}
\newcommand{\zplus}{$z^+$}
\newcommand{\zminus}{$z^-$}

\begin{document}

   \title{Polarization geometry of magnetohydrodynamic turbulence}
  
   \titlerunning{}

   \subtitle{}

   \author{Raphael Skalidis\thanks{Hubble Fellow}}

   \institute{TAPIR, Mailcode 350-17, California Institute of Technology, Pasadena, CA 91125, USA \and Faculty of Engineering and Natural Sciences, Sabanci University, Orhanli-Tuzla 34956, Istanbul, Turkey \\
   \email{rafail.skalidis@sabanciuniv.edu}
   }
   \date{Received ; accepted}

 
  \abstract
   {We introduce a geometric framework that organizes the second-order scalar statistics of the Elsasser fields into polarization states in a generalized Poincaré framework. Energy, cross-helicity, residual energy, and the phase-lag between counter propagating waves emerge as complementary parameters. We derive a Bloch-analogue equation governing the evolution of polarization states and their degree of polarization, distinguishing linear precession from nonlinear changes in polarization. We propose a phenomenological connection between polarization geometry and cascade dynamics, and identify solar wind measurements that can test whether transitions in spectral scaling accompany changes in polarization.}


   \maketitle
%

\section{Introduction}

Magnetohydrodynamic (MHD) turbulence exhibits a complexity that remains challenging to capture within a unified theoretical framework. The partition and transfer of energy between kinetic and magnetic fluctuations can depart substantially from the equipartition expected in simple descriptions of Alfvénic turbulence \citep{skalidis_tassis2021,skalidis_2021_bpos,skalidis_2023.analytical,beattie_2022_energy_balance}. Additionally, conventional models predict uniform scaling properties throughout the inertial range, either $E \propto k^{-3/2}$ \citep{Iroshnikov_1964,boldyrev_2005.dynamic.alignment,boldyrev_2006.dynamic.alignment} or $E \propto k^{-5/3}$ \citep{sridhar_1994,goldreich_1995}. Yet solar wind measurements and numerical simulations frequently reveal a richer behavior \citep{podesta_2007.spectral.exponents,podesta_2011.spectrum.cross.hel,sioulas_2023.magnetic.field.evolution,skalidis_2026.residual.energy} with a shallow $k^{-1}$ scaling range at large scale preceding the inertial range \citep{matthaeus_1986.1f.noise,wicks_2013.residual_energy.cross_helicity.energy.containing}, while in some cases distinct inertial subranges coexist \citep{mondal_2025.1/f.two.subinertial.ranges,wu_2025.solar.wind.subranges}. The physical origin of this spectral diversity, and its connection to the underlying state of the turbulent fluctuations, remain unclear.

Scalar diagnostics commonly used to characterize MHD turbulence, including cross-helicity ($H_c = \textbf{u} \cdot \textbf{b}$) and residual energy ($E_r = u^2 - b^2$), have been associated with the variations in the scaling properties of MHD turbulence \citep{chen_2016.review.solar.wind}.  Additional factors known to influence the spectral properties include background inhomogeneities \citep{velli_1989.reflection.driven.turb}, parametric instabilities \citep{chandran_2018.parametric.instability}, and the fluctuating-to-mean field ratio ($\delta B /B_0$) \citep{muller_grappin_2005.power_spectra.incompressible.MHD,brodiano_2026.mhd.sims.1/f.scaling}. Although these quantities arise or probe distinct physical mechanisms, they all refer to a common property of the turbulence: the degree of coherence among the fluctuating fields. 

If turbulence is understood as a process that decorrelates interacting modes though nonlinear phase mixing, coherence opposes this randomization by preserving the phase and amplitude relationships between those modes over extended timescales. Consequently, coherent large-scale fluctuations weaken nonlinear interactions and delay the development of the turbulent cascade \citep{wicks_2013.alignment,wicks_2013.residual_energy.cross_helicity.energy.containing}. 

Cross-helicity normalized by total energy ($\sigma_c$) measures the alignment between velocity and magnetic fluctuations: an Alfv\'en wave, where $\sigma_c=1$ is a maximally coherent state. Similarly, normalized residual energy ($\sigma_r$) quantifies the balance between kinetic and magnetic energy: when $\sigma_r=1$, the fluctuation represents a long-lived coherent structure. The interplay between coherent structures and propagating waves is central to the energy cascade \citep{zhao_2025.review.structures.waves}. Inhomogeneities and parametric instabilities naturally excite coherent fluctuations, while strong guide fields ($\delta B/B_0 < 1$) can sustain the emerging coherence over prolonged times \citep{velli_1993.reflection.driven,meyrand_2025.reflection.driven.turb}. 
In this work, we show that the aforementioned coherence metrics naturally link to the polarization properties of interacting modes. The various diagnostics are geometrically represented as distinct polarization states on the Poincar\'e sphere, allowing coherent and dispersive interactions to be described within a common mathematical framework. 
This formulation provides a natural framework for understanding the diverse processes that govern incompressible turbulence.

\section{Theory development}

Under Elsasser's variable transformation, $z^{\pm} = u \pm \delta B/\sqrt{4\pi \rho}$, the equations of incompressible MHD take the following form \citep{elssaser_1950.variables}:
\begin{equation}
	\label{eq:incomp_MHD}
	\partial_t \bm{z}^\pm \mp \left( \bm{V_A} \cdot \nabla \right  ) \bm{z}^{\pm} + \left( \bm{z}^\mp \cdot \bm{\nabla} \right) \bm{z}^\pm = - \bm{\nabla} P
\end{equation}
where $V_A = B /\sqrt{4\pi \rho}$, and $P$ is the gas pressure, ensuring that the incompressibility condition ($\nabla \cdot u = 0$) is satisfied everywhere in the fluid; volume density ($\rho$) is constant. 
 
The linearized MHD equations admit two independent Elsasser solutions, corresponding to forward- (\zplus) and backward-propagating  (\zminus)  Alfvénic fluctuations. This naturally motivates the definition of a two-component state vector, $ \Psi_k \equiv (\bm{z}_k^+,\, \bm{z}_k^-)^{\rm T}$, where $\bm{z}^\pm_k$ represents the Fourier transformed variables.

The statistical properties of the fluctuations are described by the Elsasser coherence matrix,
\begin{equation}
\label{eq:density_matrix}
J (k, k') \equiv  \Psi_{k} \Psi_{k'}^\dagger 
=
\begin{pmatrix}
 z_{k}^+ \cdot z_{k'}^{+*} &
 z_{k}^- \cdot z_{k'}^{+*}  \\
 z_{k}^+ \cdot z_{k'}^{-*}  &
 z_{k}^- \cdot z_{k'}^{-*} 
\end{pmatrix}.
\end{equation}
Henceforth, we employ the assumption of local spectral coherence $k' \rightarrow k$, where $J (k, k') = J (k)$. In this case, diagonal terms represent the energies of $\bm{z}^\pm_k$, while off-diagonal terms encode their corresponding complex correlations.

The coherence matrix can be decomposed as,
\begin{equation}
\label{eq:pauli_projection}
J_{\mathbf{k}} = \frac{1}{2}\left(S_0 \mathcal{I} + S_1 \sigma_3 + S_2 \sigma_1 + S_3 \sigma_2 \right),
\end{equation}
where $\sigma_1, \sigma_2, \sigma_3$ are Pauli's traceless matrices,
\begin{equation}
\label{eq:pauli_matrices}
\sigma_1=
\begin{pmatrix}
 0 &  1\\
 1 & 0
\end{pmatrix}, 
\sigma_2=
\begin{pmatrix}
 0 &  -i\\
 i & 0
\end{pmatrix}, 
\sigma_3=
\begin{pmatrix}
 1 &  0\\
 0 & -1
\end{pmatrix},
\end{equation}
and $S_i$ corresponds to the components of the polarization state vector ($\bm{S}_k$) resulting from $\bm{z}^+_k$ and $\bm{z}^-_k$ interactions:
\begin{align}
\label{eq:stokes_components}
\bm{S}_k
\equiv 
\begin{pmatrix}
 S_0 =  |\bm{z}_k^+|^2 + |\bm{z}_k^-|^2 
& (\text{total energy}) \\[6pt]
 S_1 =   |\bm{z}_k^+|^2 - |\bm{z}_k^-|^2  
& (\text{cross-helicity}) \\[6pt]
 S_2 = 2\,\mathrm{Re}\, \left( \bm{z}_k^+ \cdot \bm{z}_k^{-*} \right)
& (\text{residual energy}) \\[6pt]
 S_3 = 2\,\mathrm{Im}\, \left( \bm{z}_k^+ \cdot \bm{z}_k^{-*} \right)
& (\text{phase lag})
\end{pmatrix}.
\end{align}
The scalar $S_0$ represents the total energy; $S_1$ represents the Elsasser imbalance, which probes the cross-helicity; $S_2$ measures the correlation between the Elsasser variables, which probes the residual energy properties; $S_3$ encodes the in-phase correlations between the two Elssaser fields. 

\begin{figure}
   \centering
   \includegraphics[width=0.8\columnwidth]{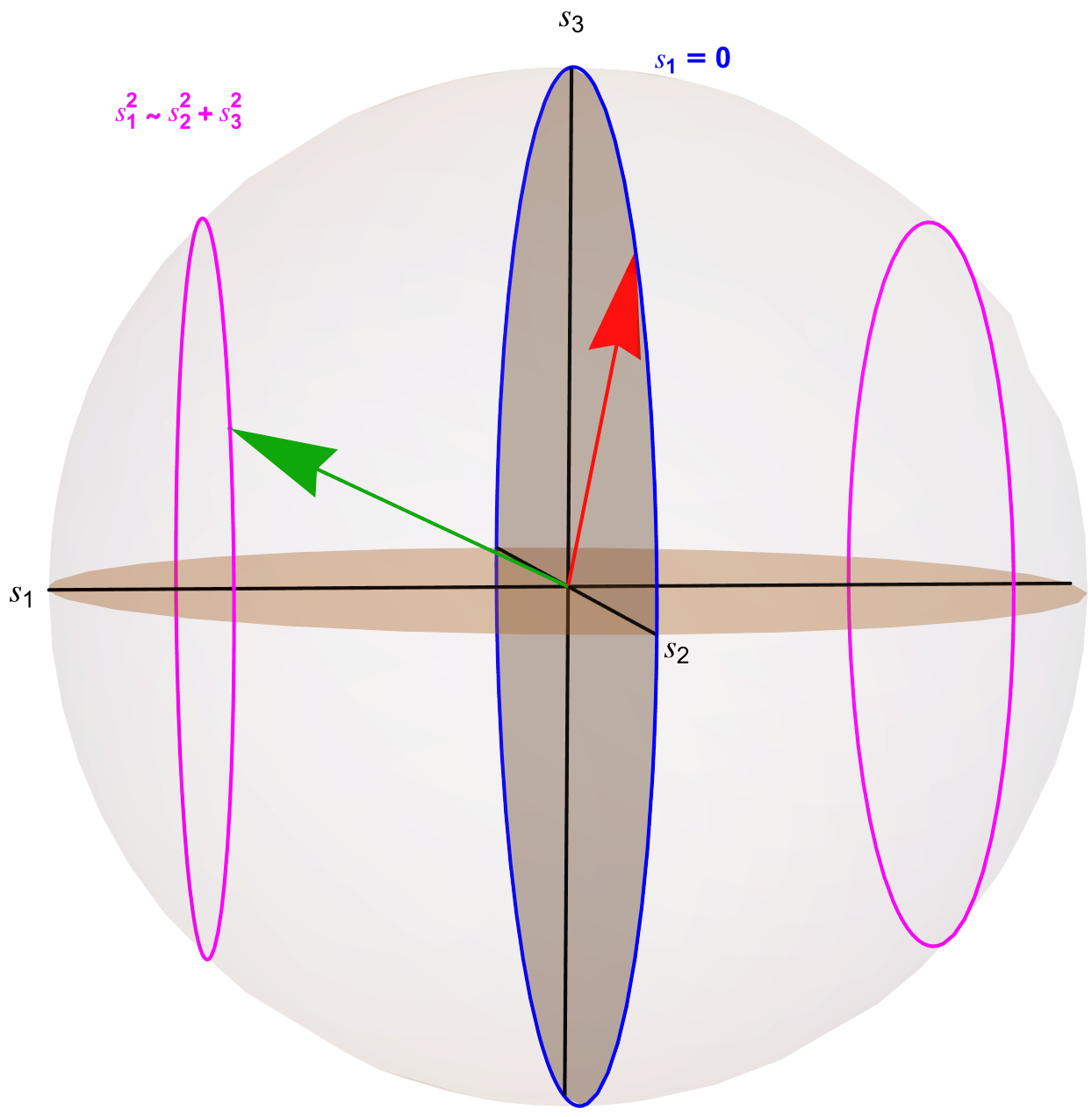}
   \caption{Geometric representation of second-order Elsasser statistics through the polarization vector $\bm{s}$ (green). In the linear limit, transverse states ($\bm{s}\perp\hat{\bm{s}}_1$) undergo coherent rotations in the $s_2$--$s_3$ plane, whereas axial states ($\bm{s}\parallel\hat{\bm{s}}_1$) have no linear precession and evolve through nonlinear coupling when present. The magenta circles mark $s_1^2\simeq s_2^2+s_3^2$, where the axial and transverse sectors have comparable magnitudes, providing a candidate geometric criterion for this transition.}
    \label{fig:poincare_sphere}
\end{figure}

The decomposition of the Elssaser coherence matrix on the Pauli basis shows that the second-order statistics of the Elsasser fields are fully specified by a coherence state vector ($\bm{S}$) under a single underlying geometry. We also notice that the Stokes representation naturally motivates the inclusion of a third scalar that is rarely treated in turbulence studies \citep[e.g.,][]{wu_yang_2024.elssaser.coherence.res.crosshel}: the phase-lag ($S_3$), which can play crucial role in turbulence dynamics \citep{nariyuki_2005.phase.coherence.alfven}. Since turbulence diagnostics are usually normalized by the energy, we focus on the normalized coherence state vector $\bm{s} = \left(S_1, S_2, S_3 \right)/S_0$. Its geometry on the Poincar\'e space is illustrated in Fig.~\ref{fig:poincare_sphere}.

The eigenvalues of the Elssaser coherence matrix are
\begin{equation}
	\lambda_\pm = \frac{1}{2} \left ( S_0 \pm \sqrt{S_1^2 + S_2^2 + S_3^2} \right),
\end{equation}
Which, upon normalization by $S_0$, become,
\begin{equation}
	\tilde{\lambda}_\pm = \frac{1 \pm p}{2}, ~~ p = |\bm s| = \frac{\sqrt{S_j S_j}}{S_0},
\end{equation}
where repeated indices imply summation and $p \in [0,1]$ measures the degree of polarization (coherence) of the turbulence state. 

The limits $p=1$ and $p=0$ correspond to pure and maximally mixed states respectively. These states represent coherent fluctuations, such as individual Alfv\'en waves ($\sigma_c = \pm1$) or coherent structures ($\sigma_r = \pm 1$), for which the turbulence state is described by a single point on the Poincar\'e sphere. In contrast, when $0 \leq p \leq 1$, the states are mixed, comprising  a superposition of fluctuations with different amplitudes and phases. 

\subsection{Dynamical evolution of the polarization states}
\label{sec:theory_evolution}

The key question we wish to address is how turbulence states evolve across the Poincar\'e space. Through the Fourier transform,  $\bm{z}_k^\pm=i\phi_k^\pm \hat{e}_k^\pm$ \citep{goldreich_1995}, we write Eq.~\ref{eq:incomp_MHD} as
\begin{equation}
	\label{eq:incomp_MHD_fourier}
	\partial_t \phi^\pm \mp i \left( \bm{k} \cdot \bm{V}_A \right  ) \phi_k^{\pm}  = \mathcal{N}_k^\pm ,
\end{equation}
where the nonlinear term is equal to
\begin{equation}
	\label{eq:nonlinear_term}
	\mathcal{N}_k^\pm = \frac{1}{(2\pi)^3} \int d^3q \int d^3p\, \phi_q^\pm \phi_p^\mp \mathcal{M}_{kqp}^{\pm} |\bm{k}|	\delta(\bm{p} + \bm{q} - \bm{k}),
\end{equation}
\begin{equation}
	\label{eq:projection}
	\mathcal{M}_{kqp}^{\pm} = (\hat{e}_k^\pm\cdot\hat{e}_q^\pm) (\hat{e}_k^\pm\cdot\hat{e}_p^\mp);
\end{equation}
$\bm{k}$, $\bm{q}$, and $\bm{p}$ correspond to the wavectors of the interacting modes. 

From Eq.~\ref{eq:incomp_MHD_fourier}, we obtain the time evolution of the Stokes vector,
\begin{equation}
\label{eq:density_matrix_time_evolution}
\frac{\partial \bm{S}}{\partial t}
=
\begin{pmatrix}
 \dot S_0 \\
 \dot S_1 \\
 \dot S_2 \\
 \dot S_3 \\
\end{pmatrix}
=
2
\begin{pmatrix}
 0 \\ 
0\\\
-\omega_k S_{3}  \\
+\omega_k S_{2}  
\end{pmatrix} + 2
\begin{pmatrix}
  \mathrm{Re}\, \left [  \Phi_k^\dagger  \mathcal{N}_k  \right]\\ 
 \mathrm{Re}\, \left [ \Phi_k^\dagger \sigma_3 \mathcal{N}_k \right]\\\
 \mathrm{Re}\  \left[ \Lambda_k \right ]\\
  \mathrm{Im}\  \left[ \Lambda_k \right ] 
\end{pmatrix},
\end{equation}
where $\omega_k = \bm{k} \cdot \bm{V}_A$, $\Phi_k^{\mathsf T}=(\phi_k^+,\phi_k^-)$, $\mathcal{N}_k^{\mathsf T}=(\mathcal{N}_k^+,\mathcal{N}_k^-)$, and
\begin{equation}
\Lambda_k = \phi_k^+\mathcal{N}_k^{-*} + \phi_k^{-*}\mathcal{N}_k^+ .
\end{equation}

The evolution of the normalized Stokes vector $\bm s$ on the Poincar\'e sphere is governed by:
\begin{equation}
	\label{eq:bloch_analogue}
	\partial_t \bm{s} = \bm{\Omega}_k \times \bm{s}_k  + \overbrace{ \bm{\mathcal{D}}_k - \bm{s} \partial_t \ln{S_0}}^{\mathcal{R}_k} \Rightarrow \partial_t \bm{s} = \bm{\Omega}_k \times \bm{s}_k  + \mathcal{R}_k,
\end{equation}
where $\bm{\Omega}_k = (2\omega_k, 0, 0) = 2k_\parallel V_A \hat{\tilde{s}}_1$ is a linear term and $\bm{\mathcal{D}}_k$ collects the nonlinear terms from Eq.~\ref{eq:density_matrix_time_evolution},
\begin{equation}
	\label{eq:non_linear_bloch}
	 \bm{\mathcal{D}}_k = \frac{2}{S_0}
	\begin{pmatrix}
 		\mathrm{Re}\, \left [ \Phi_k^\dagger \sigma_3 \mathcal{N}_k \right]\\\
 		\mathrm{Re}\  \left[ \Lambda_k \right ]\\
 		\mathrm{Im}\  \left[ \Lambda_k \right ] 
	\end{pmatrix}.
\end{equation}
Eq.~\ref{eq:bloch_analogue} probes how the coherence vector evolves in the Poincar\'e space and is reminiscent of the Bloch equation, which describes the behavior of nuclear magnetization vectors in quantum mechanics. 

The evolution of the degree of polarization follows directly by taking the scalar product of Eq.~\ref{eq:bloch_analogue} with $\bm{s}$,
\begin{equation}
	\label{eq:pol_degree_evolution}
	\partial_t p^2=2\bm{s}\cdot\bm{\mathcal{R}}_k
	=2\bm{s}\cdot\bm{\mathcal{D}}_k-2p^2\partial_t\ln S_0.
\end{equation}
Changes in $p$ are governed by the radial contribution of the nonlinear terms relative to the accompanying energy variation. Nonlinear terms can either enhance or suppress $p$, while the linear rotation term preserves $p$ because $\bm{s}\cdot(\bm{\Omega}_k\times\bm{s})=0$ (Appendix~\ref{sec:appendix_traj}).

\subsection{Dependence of the polarization dynamics on the initial state}
\label{sec:initial_state}

Because the generalized Poincar\'e sphere obeys SU(2)-like algebra, rotation between different polarization states are non-commuting. Consequently, turbulence interactions depend not only on the instantaneous polarization state, but also on the path through which the states evolve in the Poincar\'e space. Different initial polarization states give rise to distinct trajectories on the Poincar\'e sphere (Appendix~\ref{sec:appendix_traj}), which in turn could influence the properties of the resulting turbulent cascade. 

Initial states along the $s_1$ axis evolve nonlinearly, $\partial_t s_k \sim \mathcal{R}_k$, as a consequence of their geometry, for which $\bm{s} \parallel \hat{s}_1 \Rightarrow \bm{\Omega} \times \bm{s} = 0$ (e.g., Fig.~\ref{fig:perp_sector}). For finite interacting populations, $\mathcal{N}^\pm\sim k_\perp\delta z^\mp\delta z^\pm$ gives the usual estimate $\tau_{\rm nl}^\pm\sim(k_\perp\delta z^\mp)^{-1}$. For comparable amplitudes, $\bm{s}$ is dimensionless and $\mathcal{R}\sim k_\perp\delta z$ is a characteristic rate, giving $\tau_s\sim(k_\perp\delta z)^{-1}$ for order-unity nonlinear changes.

Contrary to $s_1$, the evolution of perpendicular fluctuations, $s_2$ and $s_3$, are to first order governed by the linear rotation term in Eq.~\ref{eq:bloch_analogue} when $\bm{s}_k \perp \hat{s}_1$ (e.g., Fig.~\ref{fig:parallel_sector}). Because $\bm{\Omega} \times \bm{s} \neq 0$, the polarization state vector rotates coherently in the $s_2 - s_3$ plane with a characteristic timescale $\partial_t s \sim s k_\parallel V_A \Rightarrow \tau_s \sim 1/\left( k_\parallel V_A \right)$.

In the linear regime, the radius of linear rotations is conserved, 
\begin{equation}
	\label{eq:conservation_relation}
	S_2^2 + S_3^2 = |\bm{z}^+_k \cdot \bm{z}^{-*}_k| \sim \rm{constant}.
\end{equation}
For the normalized Stokes parameters with the closure enforced by Eq.~\ref{eq:conservation_relation}, we obtain,
\begin{align}
	\label{eq:s2_analytical_sol}
	s_2 (t) \propto \cos \left( 2 \omega t + \theta_0 \right) \\
	\label{eq:s3_analytical_sol}
	s_3 (t) \propto \sin  \left( 2 \omega t + \theta_0 \right) 
\end{align} 
where $\theta_0$ is an initial phase.		 

Linear rotations in the $s_2 - s_3$ plane differ from the classical picture of Alfv\'enic interactions. In the standard framework, $z^+$ and $z^-$ represent independent modes whose interactions drives the cascade \citep{goldreich_1995}. The Elsasser coherence matrix (Eq.~\ref{eq:density_matrix}) encodes this independence: when $z^+$ and $z^-$ are uncorrelated, the matrix is diagonal. However, when the coupling between the Elssaser variables becomes significant $\bm{z^+} \cdot \bm{z^-}$, $ \rm{Tr(J_k)} \sim |\bm{z}^+_k \cdot \bm{z}^{-*}_k| \sim S_0$, underscoring that the Elsasser fields are no longer independent.

The normalized eigenvalues of these states are $\tilde{\lambda}_+ \approx 1$ and $\tilde{\lambda}_- \approx 0$, corresponding to a nearly fully polarized state with $p \approx 1$. In the fully polarized limit, the smaller eigenvalue vanishes and the coherence matrix becomes rank one, so that a single eigenvector specifies the polarization state. This eigenvector is a coherent superposition $z^+$ and $z^-$,
\begin{equation}
	\tilde{z}^{\pm}_k = \frac{1}{\sqrt{2}}\left[z^+_k \pm e^{i\theta_k(t)}z^-k\right], \qquad \theta_k(t)=\theta_{k,0}+2\omega_k t.
\end{equation} 

In this linear limit, the polarization vector undergoes pure precession about the $\hat{s}_1$ axis with angular frequency $2|\omega_k|$. Consequently, the $s_2$ and $s_3$ components oscillate with a phase difference of $\pi/2$ (Eqs.~\ref{eq:s2_analytical_sol}, \ref{eq:s3_analytical_sol}), describing a reversible exchange between kinetic ($u^2$) and magnetic ($b^2$) energy at fixed wavevector. When nonlinear decorrelation is sufficiently weak, these states can persist as coherent wave motions, providing a natural geometric description of phase-coherent counter-propagating Alfv\'enic fluctuations, including standing waves \citep{de_giorgio_2017}.

The linear rotation term in Eq.~\ref{eq:bloch_analogue} would eventually become comparable to the nonlinear term. At this point, the linear and nonlinear contributions satisfy $\Omega_k s_k \sim \mathcal{R}_k$, recovering the critical balance condition \citep{goldreich_1995},
\begin{equation}
	\label{eq:critical_balance}
	k_\parallel V_A \sim k_\perp \delta V_A.
\end{equation}
Within the present framework, critical balance marks the onset of the transition from (linear) coherent to mixed (nonlinear) polarization dynamics.

Nonlinear interactions can alter a polarized state by either reducing its degree of polarization or reorienting it toward the imbalance ($\hat{s}_1$) axis. The first process corresponds to a loss of coherence, while the second transfers coherence from transverse Elsasser correlations (\(s_2,s_3\)) to $s_1$, corresponding to a transition from a balanced to an imbalanced state. The transition between these regimes occurs when the parallel and perpendicular polarization sectors are comparable in magnitude
\begin{equation}
	\label{eq:transition_scale_condition}
	s_\parallel^2 \sim s_\perp^2 \Rightarrow s_1^2 \sim s_2^2 + s_3^2 \Rightarrow 
	|\bm{u}\cdot\bm{b}| \sim |\bm{z}_k^+ \cdot \bm{z}_k^{-*}|.
\end{equation}

Geometrically, this condition is reached when the polarization vector lies midway between the imbalance axis and the \(s_2\)--\(s_3\) plane (magenta circles in Fig.~\ref{fig:poincare_sphere}). In the limit \(|s_3| \ll |s_{1,2}|\), Eq.~\ref{eq:transition_scale_condition} reduces approximately to \(|\sigma_c| \sim |\sigma_r|\). The two evolutionary pathways, depolarization and migration, are not mutually exclusive and may operate simultaneously.

\begin{figure}
   \centering
   \includegraphics[width=\columnwidth]{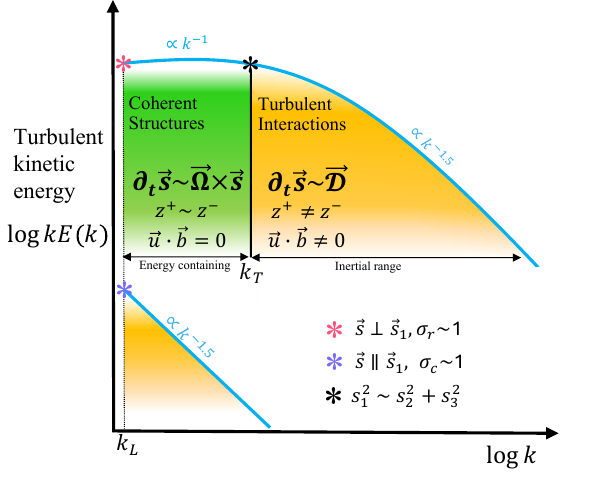}
    \caption{Different initial polarization geometries define distinct evolutionary paths for turbulence dynamics. A migration of the $\bm{s}$ vector from the $s_2$ -- $s_3$ plane toward $\hat{s}_1$ is expected to be associated with a scaling transition change. At the transition scale, $k_{T}$, $s_1^2 \sim s_2^2 + s_3^2$. In contrast, $s_1$-dominated states are governed primarily by nonlinear interactions and are associated with steeper scalings.
    }
    \label{fig:phenomenology}
\end{figure}

\section{Measurable consequences and tests} 
\subsection{Polarization diagnostics in the solar wind}

The degree of polarization provides a unified framework for diagnostics already linked to solar-wind scaling and alignment \citep{wicks_2013.alignment,wicks_2013.residual_energy.cross_helicity.energy.containing}. In a predominantly Alfvénic fast stream, \citet{wicks_2013.residual_energy.cross_helicity.energy.containing} found steep structure functions extending to low frequencies for nearly balanced fluctuations with comparable kinetic and magnetic energies ($\sigma_c,\sigma_r) \approx (0, 0)$, whereas magnetically dominated or strongly imbalanced populations ($|\sigma_c| \approx 1$ or $|\sigma_r| \approx 1$) exhibited broader $1/f$ ranges. Our geometric framework suggests that small values of $\sigma_c$ and $\sigma_r$ do not necessarily imply weak polarization. Substantial polarization can still be present when $s_3$ is significant.

The third coordinate ($s_3$) therefore offers an additional observational test. Solar wind intervals matched in $\sigma_c$ and $\sigma_r$ can differ in $s_3$ and potentially in their interaction times. Comparing their spectra would test whether this phase information contributes to the scaling differences. 

Because the Elsasser coherence matrix is Hermitian, its eigenvalues must remain positive, imposing the following geometric constraint,
\begin{equation}
	\label{eq:geometric_constraint}
	\langle s_1^2\rangle+\langle s_2^2\rangle+\langle s_3^2\rangle\leq1.
\end{equation}
For a population of normalized states, let $\mu_i=\langle s_i\rangle$ and $\delta s_i^2=\langle(s_i-\mu_i)^2\rangle$, we obtain that
\begin{equation}
	\label{eq:spread_constraint}
	\delta s_1^2+\delta s_2^2
	=\langle p^2\rangle-\sum_{i=1}^3\mu_i^2-\delta s_3^2
	\leq1-\sum_{i=1}^3\mu_i^2-\delta s_3^2.
\end{equation}
The variance in $s_3$ constrains the variance available to cross-helicity and residual energy. Solar wind measurements reveal radial evolution in their joint distribution, characterized by decreasing Alfvénic imbalance and an increased prevalence of magnetically dominated, low-cross-helicity states \citep{sioulas_2023.magnetic.field.evolution}. Within our framework, these trends can be tested by comparing the scale-dependent variances of the Stokes components at fixed scales, thereby assessing whether the observed evolution is accompanied by a reduction in phase coherence. Measurements of $\langle s_3 \rangle$, $\delta s_3^2$, and $\langle p^2\rangle$ would provide the missing information needed for a complete comparison.

\subsection{Polarization geometry and the turbulent cascade}

Our physical proposal is that polarization geometry helps organize different cascade regimes through the persistence and redistribution of amplitude and phase correlations. The dependence of turbulent spectra on the driving mechanism, even at comparable global Mach numbers, suggests that the polarization geometry may influence the cascade dynamics \citep{skalidis_2026.residual.energy}. 

Figure~\ref{fig:phenomenology} illustrates a phenomenological connection between polarization dynamics and inertial range scaling. Large-scale inhomogeneities can generate correlated Elsasser populations, providing a route to $k^{-1}$ spectra \citep{velli_1989.reflection.driven.turb,velli_1993.reflection.driven}. Dynamic alignment motivates a possible $k^{-3/2}$ scaling \citep{boldyrev_2005.dynamic.alignment,boldyrev_2006.dynamic.alignment,perez_boldyrev_2009.cross.helicity.role}, while phase-mixed populations are expected to approach $k^{-5/3}$ scalings \citep{goldreich_1995}. 

The proposed links between polarization geometry and spectral properties depend on both the polarization geometry and the nonlinear dynamics. Axial polarization states have no linear precession and can evolve nonlinearly. Conversely, linear precession changes the kinetic–magnetic energy partition at fixed wavevector, without transferring energy between scales.

This framework makes three testable predictions. First, spectral transitions should accompany systematic changes in polarization, with $s_3$ providing information beyond the conventional $\sigma_c$ and $\sigma_r$. Second, populations evolving from transverse coherence toward imbalance should preferentially exhibit breaks near the proposed condition $s_1^2\sim s_2^2+s_3^2$ (Eq.~\ref{eq:transition_scale_condition}). Third, stronger guide fields may favor persistent coherent populations and consequently more prominent shallow spectral ranges, a hypothesis supported by the simulations of \citet{brodiano_2026.mhd.sims.1/f.scaling}.

The framework provides a geometric language for characterizing coherent interactions and their potential impact on the turbulent cascade. It is formally valid for incompressible MHD turbulence, while inhomogeneities introduce additional dynamical terms that lie beyond the scope of the present treatment \citep{magyar_2019.elssaser.compressible.MHD,Banerjee_2013}.

\bibliographystyle{aa}
\bibliography{bibliography}
 
\begin{acknowledgements}
I thank the anonymous referee for their valuable suggestions. I am especially grateful to K. Tassis and P. F. Hopkins for insightful discussions and their continued encouragement throughout the development of these ideas. I also thank J. R. Beattie, A. Tritsis,  and N. Sioulas for their constructive feedback. Support for this work was provided by NASA through the NASA Hubble Fellowship grant \#~HST-HF2-51566.001 awarded by the Space Telescope Science Institute, which is operated by the Association of Universities for Research in Astronomy, Inc., for NASA, under contract NAS5-26555. A supplementary Mathematica notebook with interactive figures is archived on Zenodo
\citep{skalidis_2026.bloch.notebook}.
\end{acknowledgements}

\begin{appendix}

\section{Representative trajectories}
\label{sec:appendix_traj}

We visualize examples of trajectories obeying the Bloch-analogue equation
(Eq.~\ref{eq:bloch_analogue}) on the Poincar\'e sphere. We employ a simple
dissipative toy model for the nonlinear term,
$\bm{\mathcal{R}_k} \rightarrow -\boldsymbol{\Gamma}\bm{s}$, such that the
Bloch-analogue equation becomes
\begin{equation}
	\partial_t \bm{s}
	=
	\bm{\Omega}_k \times \bm{s}
	-
	\boldsymbol{\Gamma}\bm{s},
\end{equation}
where $\boldsymbol{\Gamma}$ represents a symmetric dissipation tensor,
\begin{equation}
\boldsymbol{\Gamma} =
\begin{pmatrix}
\gamma_{11} & \gamma_{12} & \gamma_{13} \\
\gamma_{12} & \gamma_{22} & \gamma_{23} \\
\gamma_{13} & \gamma_{23} & \gamma_{33}
\end{pmatrix}.
\end{equation}
The diagonal coefficients $\gamma_{ij}$ control the damping rates along the three polarization axes, associated with cross-helicity, residual energy, and phase coherence, while the off-diagonal coefficients describe coupling between these polarization components. Allowing different damping rates accounts for the possibility that nonlinear interactions preferentially destroy some polarization properties more efficiently than others.

\begin{figure}
\centering
  \includegraphics[width=0.7\columnwidth]{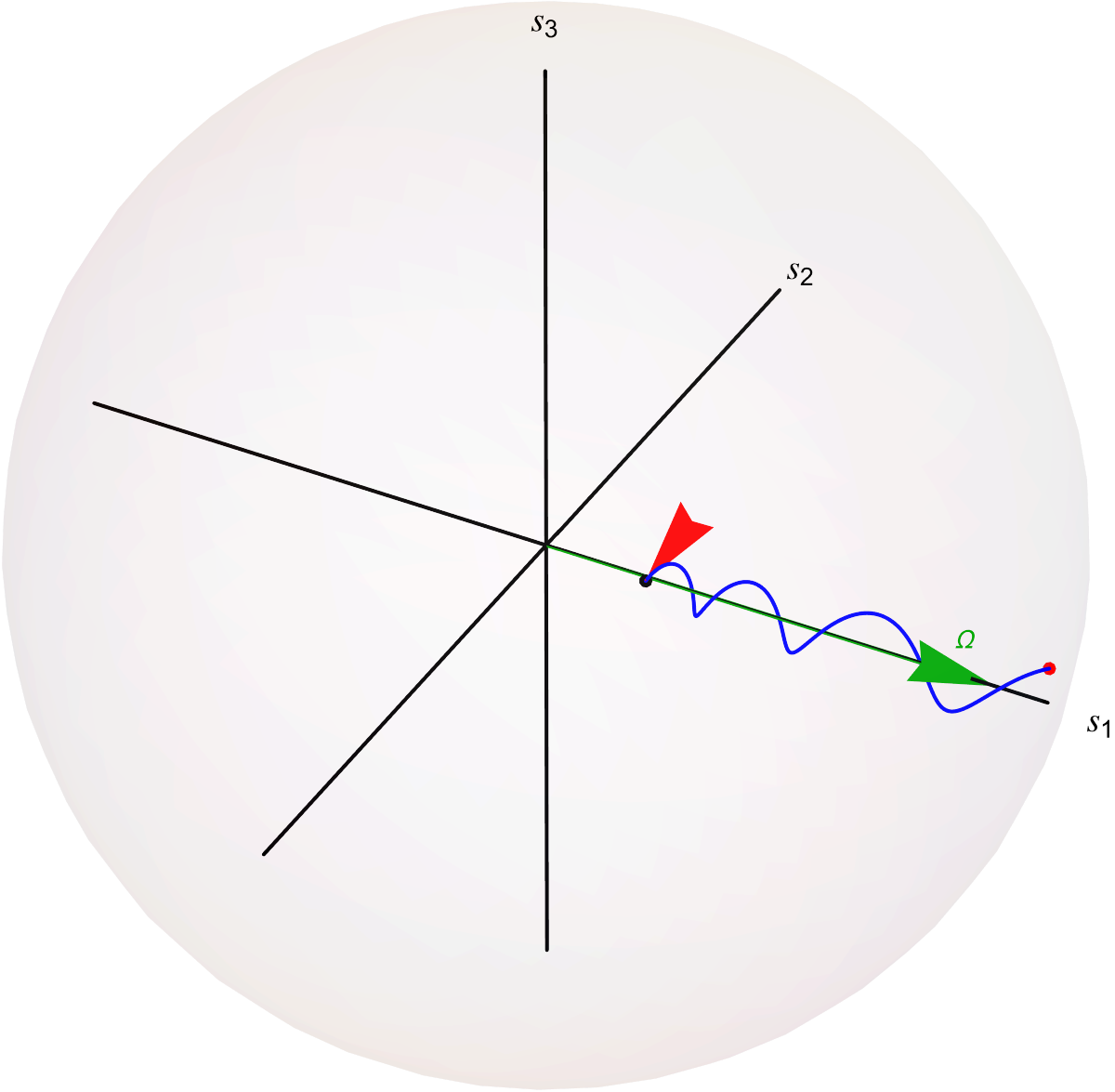}
\caption{An illustrative trajectory of a dissipating polarization vector initialized parallel to the $\hat{s}_1$ axis, with $\gamma_{11}=\gamma_{22}=0.15$, $\theta_0=\pi/2.1$, $\phi_0=0$, and $p=1$. All other $\gamma_{ij}$ are set to zero. The red and black points indicate the initial and final positions of the trajectory, respectively.}
\label{fig:perp_sector}
\end{figure}

Fig.~\ref{fig:perp_sector} illustrates a characteristic trajectory for a mode initialized along the cross-helicity axis, $\hat{s}_1$. Here, $\theta_0$ and $\phi_0$ are the initial polar and azimuthal angles of the polarization vector, respectively. In this case, the trajectory is dominated by the dissipative term, leading to a rapid loss of coherence. The linear rotation term causes the trajectory to spiral around the cross-helicity axis. In the limit of sufficiently slow rotation, the trajectory approaches a nearly radial decay.

\begin{figure}
\centering
 \includegraphics[width=0.7\columnwidth]{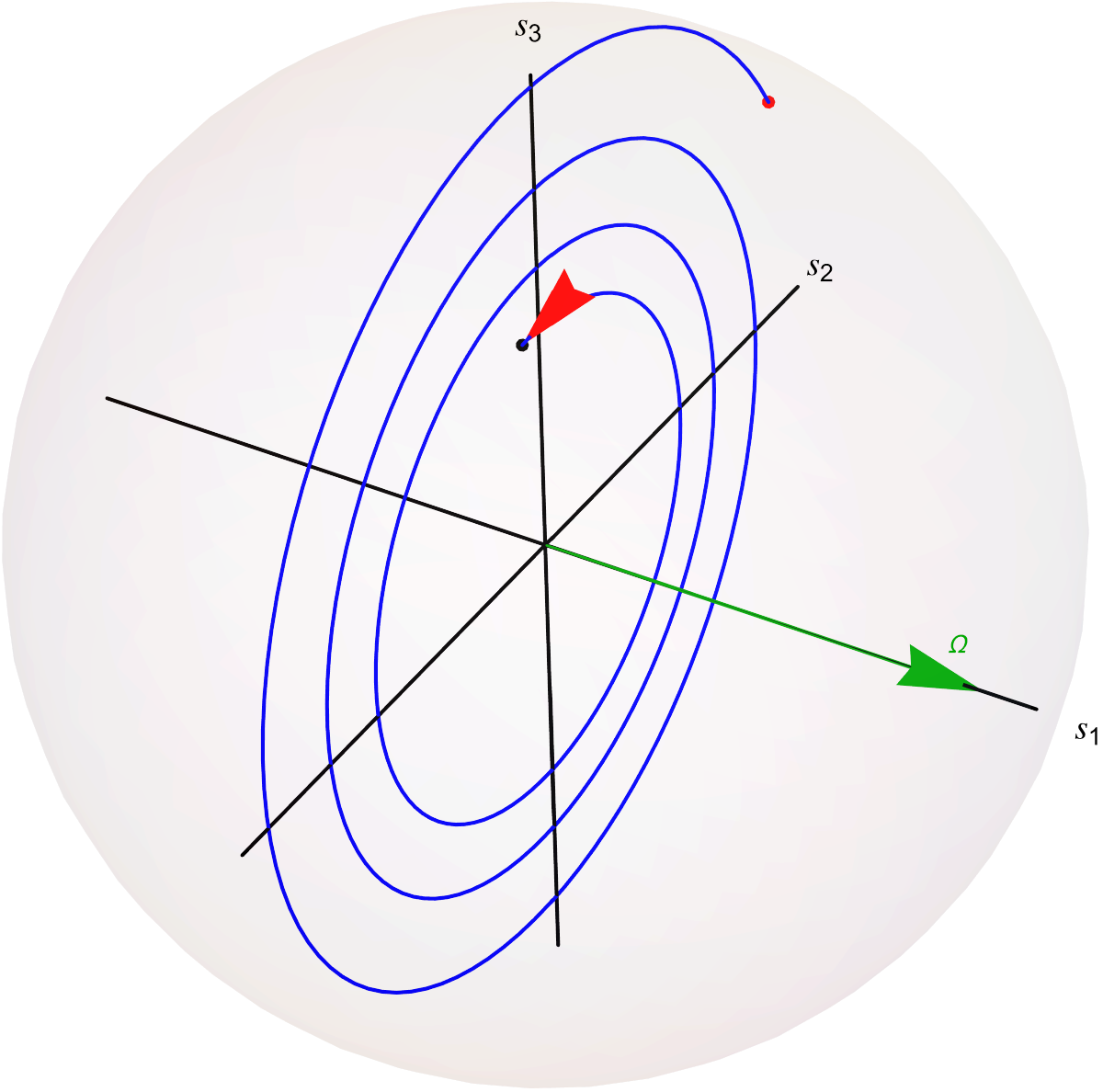}
\caption{An illustrative trajectory of a dissipating polarization vector, initialized in the $s_2$--$s_3$ plane, with $\gamma_{11}=\gamma_{22}=0.15$, $\theta_0=\pi/3$, $\phi_0=\pi/2$, and $p=1$. All other $\gamma_{ij}$ are set to zero. The red and black points indicate the initial and final positions of the trajectory, respectively.}
\label{fig:parallel_sector}
\end{figure}

Fig.~\ref{fig:parallel_sector} illustrates a characteristic trajectory
on the Poincar\'e sphere for a mode initialized in the $s_2-s_3$ plane.
 The trajectory is dominated
by the linear rotation term, while its radius decreases with time as a
consequence of the dissipative term, indicating a loss of polarization
coherence. 

\begin{figure}
\centering
 \includegraphics[width=0.8\columnwidth]{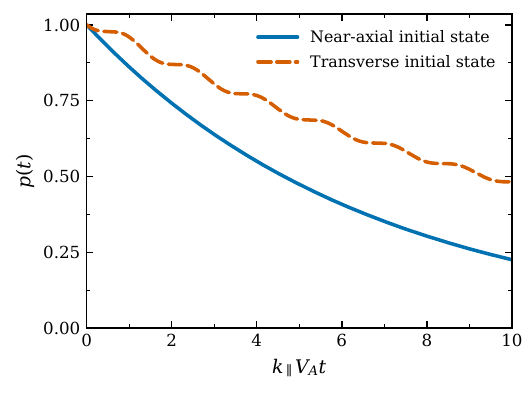}
\caption{Evolution of the degree of polarization $p(t)=|\bm{s}(t)|$ for the trajectories shown in Figs.~\ref{fig:perp_sector} and~\ref{fig:parallel_sector}. The blue solid curve corresponds to the initially near-axial state and the orange dashed curve to the initially transverse state. Both have $p(t=0)=1$ and evolve with $\boldsymbol{\Omega}_k=2k_\parallel V_A\hat{s}_1$ and $\boldsymbol{\Gamma}=\mathrm{diag}(\gamma,\gamma,0)$, where $\gamma/(k_\parallel V_A)=0.15$. Time is expressed in units of $(k_\parallel V_A)^{-1}$.}
\label{fig:polarization_degree}
\end{figure}

The corresponding evolution of the degree of polarization is shown in Fig.~\ref{fig:polarization_degree}. For the adopted model, Eq.~\ref{eq:pol_degree_evolution} implies
\begin{equation}
    \partial_t p^2
    = -2\bm{s}\cdot\boldsymbol{\Gamma}\bm{s}
    = -2\gamma(s_1^2+s_2^2).
\end{equation}
Both states depolarize monotonically. The initially near-axial state decays more rapidly, whereas the transverse state periodically experiences weaker damping as precession rotates the polarization vector toward the undamped $s_3$ direction. Linear precession does not itself change $p$, but it changes the instantaneous damping rate by reorienting the vector relative to the anisotropic dissipation tensor. This contrast illustrates the dependence of depolarization on the assumed polarization state. An interactive Mathematica notebook allows readers to explore the polarization trajectories and the evolution of the degree of polarization under the assumed toy model\footnote{\url{https://www.wolframcloud.com/obj/e62da2cb-ec9c-4a1b-9bfa-ee811c9599b1}}.

\end{appendix}

\end{document}